\documentclass[dvipdfmx]{article}
\usepackage{tikz}
\begin{document} 

\begin{center}

{\large 
The Free Energy of the Square Lattice Ising Model \\
with Interactions Alternating in Horizontal and Vertical Directions. 
}
\vspace{0.6cm}

Kazuhiko Minami
\vspace{0.6cm}

Graduate School of Mathematics, Nagoya University, \\
Furo-cho, Chikusa-ku, Nagoya, Aichi, 464-8602, JAPAN.

\end{center}

\begin{abstract}
The free energy of the Ising model on the square lattice 
with alternating interactions in both horizontal and vertical directions is exactly derived. 
This model is distinct from the checkerboard Ising model. 
The result includes Onsager's free energy as a special case, 
and also includes Lee-Yang's  free energy with an imaginary field, 
and relates these two solutions via continuous parameters. 
The result includes a generalization of  Lee-Yang's result  
to cases with four different couplings. 
It is also derived that 
each imaginary magnetic field $i\pi/2$ applied to a lattice site 
corresponds to a single frustrated square in its dual lattice. 

\end{abstract}


\noindent
e-mail: minami@math.nagoya-u.ac.jp

\section{Introduction}

Ising models have a simple and basic structure, 
are in many cases exactly solvable, 
and many of them show critical phenomena.
The square lattice Ising model was solved by Onsager\cite{44Onsager}, 
and the free energy was re-derived in a simpler manner by Kaufman\cite{49Kaufman}. 
Since then, numerous derivations of the free energy for this model have been obtained,   
based on the transfer matrix method, 
using fermionization techniques, Yang-Baxter type commutation relations, 
or based on the combinatorial methods, etc.
\cite{14Minami}

The free energy of the triangular lattice Ising model 
has been derived by many authors. 
This result naturally yields the free energy of the honeycomb lattice Ising model, 
because they are mapped to each other through a dual transformation. 
The free energy for the isotropic case was obtained by 
Husimi and Syozi\cite{50HusimiSyozi-50Syozi}, and 
Wannier\cite{50Wannier-50WannierErr}. 
The free energy for general interactions was obtained by 
Temperley\cite{50Temperley}, 
Houtappel\cite{50Houtappel}, and 
Newell\cite{50Newell}. 
All these results were derived using the transfer matrix method. 
Potts\cite{55Potts} obtained the free energy of the triangular case 
using the combinatorial method of Kac and Ward. 
Wolff and Zittartz\cite{82Wolff} 
obtained the free energy on triangular and honeycomb lattices 
with interactions of layered structures. 
Wu and Hu\cite{02Wu} calculated the partition functions with various boundary conditions 
using the Grassmann path integral approach.

The free energies of the checkerboard lattice Ising models 
have also been obtained. 
Utiyama\cite{51Utiyama} showed the free energy of the Ising model 
on a general checkerboard-type lattice. 
Giacomini\cite{85Giacomini} obtained the exact solution of an Ising model 
with crossing and four-spin interactions on the checkerboard lattice 
through a mapping to an eight-vertex model 
that satisfies the free fermion condition.\cite{70FanWu}
R. J. Baxter \cite{86Baxter84Baxter} showed that 
in the thermodynamic limit, 
the partition function of the checkerboard Ising model can be decomposed 
into the product of four square lattice Ising model partition functions.
He also showed that 
the checkerboard Ising model is equivalent to the free fermion eight-vertex model. 

Lee and Yang\cite{52LeeYang} showed 
the free energy of the square lattice Ising model in an imaginary field $H=i\pi/2$. 
This free energy was derived by 
G. Baxter\cite{65Baxter} through a diagrammatic method, 
derived by McCoy and Wu\cite{67McCoyWu} using an expression by Pfaffian, 
derived by Gaaff\cite{74Gaaff} through a mapping to a 16-vertex model, 
and derived by Merlini\cite{74Merlini} with the use of the method 
by Kadanoff and Ceva.\cite{71KadanoffCeva} 
Marshall\cite{71Marshall} showed that 
the obtained free energy is correct only if the system size is even. 
AU-Yang and Perk\cite{84AUYangPerk} pointed out that 
the dual of the Lee-Yang system is the fully-frustrated square-lattice Ising model, 
and reduces to the dimer problem. 
Wu\cite{86Wu} considered the Ising model 
on a checkerboard lattice with crossing and four-spin interactions, 
and solved it with an imaginary field $H=i\pi/2$, 
though the system remains unsolved in zero magnetic field $H=0$. 
Lin\cite{88Lin}, and Lin and Wu\cite{88LinWu} derived the free energy with $H=i\pi/2$ 
for various two-dimensional lattices. 
Suzuki\cite{91Suzuki} showed that the dual of the Lee-Yang free energy 
yields the free energy of the Villain model\cite{77Villain}, 
which is a model of a random frustrated system. 

In this paper, 
the free energy of the Ising model on the square lattice 
with interactions alternating in two directions  is derived. 
This model is distinct from the checkerboard type Ising models, 
which have been investigated in detail as summarized above.  
The obtained free energy provides a general integral form 
that has not been written down. 
This is the main result of this paper, 
and the formula is given in section 2. 
This result includes Onsager's free energy as a special case, 
as shown in section 3, 
and also includes Lee-Yang's  free energy with an imaginary field $H=i\pi/2$ 
as a special case through the dual transformation, 
as shown in section 5. 
Therefore, it can be said that 
the result relates these two free energies via continuous parameters. 
It is also derived in section 4 that 
each imaginary magnetic field $i\pi/2$ applied to a lattice site 
corresponds to a single frustrated square in its dual lattice.  
Hence, we find that 
the distribution of the frustration 
determines the distribution of the field $i\pi/2$, 
and determines the free energy. 
It should be noted that 
this system with one interaction being negative 
corresponds to a generalization of  Lee-Yang's result  
to the cases with four different couplings, 
as can be seen from section 4 and section 5. 
In this paper, Kaufman's method is basically used,  
because this method seems to be one of the simplest 
in the calculations using the transfer matrix, though our calculations are performed in the momentum space.

\section{Formulation and the Free Energy}
Let us consider the two-dimensional Ising model 
with an alternating structure in two directions. 
The Hamiltonian is  
\begin{eqnarray}
-\beta{\cal H}
&=&
\sum_{i=1}^{N/2}
\sum_{j=1}^{M}
(K _{1}\sigma_{2i-1\, j}^z\,\sigma_{2i-1\, j+1}^z
+K _{2}\sigma_{2i\, j}^z\,\sigma_{2i\, j+1}^z)
\nonumber
\\
&+&
\sum_{i=1}^{N}
\sum_{j=1}^{M/2}
(L _{1}\sigma_{i\, 2j-1}^z\,\sigma_{i+1\, 2j-1}^z
+L _{2}\sigma_{i\, 2j}^z\,\sigma_{i+1\, 2j}^z),
\label{Hamiltonian}
\end{eqnarray}
where the periodic boundary conditions are assumed 
in both horizontal and vertical directions. 
Let us assume that $M$ and $N/2$ are even.

The partition function of this model can be written using the transfer matrix as 
\begin{eqnarray}
Z={\rm tr}\Big(V_{1}V_{22}V_{1}V_{21}\Big)^{M/2},
\end{eqnarray}
where  
\begin{eqnarray}
V_{1}
&=&
\Big(\frac{e^{K_{1}}}{\cosh K_{1}^*}\Big)^{N/2}
\Big(\frac{e^{K_{2}}}{\cosh K_{2}^*}\Big)^{N/2}
\exp(K_{1}^*\sum_{i=1}^{N/2}\sigma_{2i-1}^x)
\exp(K_{2}^*\sum_{i=1}^{N/2}\sigma_{2i}^x),
\nonumber
\\
V_{2\:j}
&=&
\exp(L_{j}\sum_{i=1}^N\sigma_{i}^z\sigma_{i+1}^z),
\label{TransferMatrix}
\end{eqnarray}
and the dual interaction $K_{i}^*$ is introduced 
for each $i=1, 2$ 
by the relation $e^{-2K_{i}}=\tanh K_{i}^*$.

Then let us introduce a series of operators: 
$\eta_{1}=\sigma_{1}^x$,
$\eta_{2}=\sigma_{1}^z\sigma_{2}^z$,
$\eta_{3}=\sigma_{2}^x$,
$\eta_{4}=\sigma_{2}^z\sigma_{3}^z$,
and generally 
$\eta_{2j-1}=\sigma_j^x$ and 
$\eta_{2j}=\sigma_{j}^z\sigma_{j+1}^z$ 
$(j=1, 2, \ldots, N)$.
These operators $\eta_j$ satisfy the commutation relations 
\begin{eqnarray}
\eta_{j}\eta_{k}
=\eta_{k}\eta_{j}
\hspace{0.4cm}
(|j-k|\geq 2),
\hspace{0.6cm}
\eta_{j}\eta_{k}
=-\eta_{k}\eta_{j}
\hspace{0.4cm}
(k=j\pm 1),
\hspace{0.4cm}
\eta_j^2=1.
\label{cond}
\end{eqnarray}
The periodic boundary condition yields 
$\eta_{1}\eta_{2N}=-\eta_{2N}\eta_{1}$. 

Next let us introduce a transformation\cite{16Minami}
\begin{eqnarray}
\varphi_j
=
\frac{1}{\sqrt{2}}
e^{i\frac{\pi}{2}(j-1)}
\eta_0
\eta_1
\eta_2
\cdots
\eta_j
\hspace{0.6cm}
(j\geq 0). 
\label{transmain}
\end{eqnarray}
Here an initial operator $\eta_0$ is introduced 
which satisfies 
$\eta_0^2=-1$, 
$\eta_1\eta_0=-\eta_0\eta_1$, 
$\eta_{2N}\eta_0=\eta_0\eta_{2N}$, and 
$\eta_k\eta_0=\eta_0\eta_k\:\:(k\neq 1, 2N)$. 
The transformation (\ref{transmain}) was introduced for general operators 
that satisfy (\ref{cond}). 
In our case, 
$\eta_{2j-1}=\sigma_j^x$ and 
$\eta_{2j}=\sigma_{j}^z\sigma_{j+1}^z$, 
let $\eta_0=i\sigma_1^z$ then we find 
$\displaystyle \varphi_0=\frac{1}{\sqrt{2}}\sigma_1^z$ 
and 
\begin{eqnarray}
\varphi_{2j}
&=&
\frac{1}{\sqrt{2}}e^{i\frac{\pi}{2}(2j-1)}
\eta_0\eta_1\eta_2\cdots\eta_{2j}
=
\frac{1}{\sqrt{2}}(\prod_{k=1}^{j}\sigma_k^x)\:\sigma_{j+1}^z,
\nonumber
\\
\varphi_{2j+1}
&=&
\frac{1}{\sqrt{2}}e^{i\frac{\pi}{2}2j}
\eta_0\eta_1\eta_2\cdots\eta_{2j+1}
=
\frac{-1}{\sqrt{2}}(\prod_{k=1}^{j}\sigma_k^x)\:\sigma_{j+1}^y,
\label{trans}
\end{eqnarray}
where we assume $\prod_{k=1}^{0}\sigma_k^x=1$. 
The transformation (\ref{trans}) is nothing but the Jordan-Wigner transformation. 
With the use of (\ref{cond}), 
it can be derived that these transformed operators $\varphi_j$ satisfy 
\begin{eqnarray}
\{\varphi_j, \varphi_k\}
=
\varphi_j\varphi_k+\varphi_k\varphi_j
=
\delta_{jk}
\label{anticom}
\end{eqnarray}
for $0\leq j, k\leq 2N-1$. 
We also find 
\begin{eqnarray}
\varphi_{j}\varphi_{j+1}
=
\frac{i}{2}
\eta_{j+1},
\hspace{0.6cm}
\varphi_{2N-1}\varphi_{0}
=
\frac{i}{2}
(-\sigma_1^x\cdots\sigma_N^x)\sigma_N^z\sigma_1^z.
\label{varphivarphieta}
\end{eqnarray}
Thus the periodic boundary condition is preserved 
if we assume the following condition in each subspace that 
\begin{eqnarray}
\varphi_{2N}
=
\left\{
\begin{array}{cl}
-\varphi_{0} & \hspace{0.3cm}\sigma_1^x\cdots\sigma_N^x=+1\\
+\varphi_{0} & \hspace{0.3cm}\sigma_1^x\cdots\sigma_N^x=-1.
\end{array}
\right.
\label{bcond}
\end{eqnarray}

Corresponding to the alternate structure of the Hamiltonian in the horizontal direction, 
let us introduce the notations 
\begin{eqnarray}
\varphi_{4j-4}=\varphi_1(j), 
\hspace{0.3cm}
\varphi_{4j-3}=\varphi_2(j), 
\hspace{0.3cm}
\varphi_{4j-2}=\varphi_3(j), 
\hspace{0.3cm}
\varphi_{4j-1}=\varphi_4(j)
\hspace{0.6cm}
(j=1, 2, \ldots, \frac{N}{2}).
\end{eqnarray}
Then we find 
\begin{eqnarray}
\varphi_{1}(j)\varphi_{2}(j)
=
\frac{i}{2}
\sigma^x_{2j-1},
\hspace{0.3cm}
\varphi_{2}(j)\varphi_{3}(j)
=
\frac{i}{2}
\sigma^z_{2j-1}\sigma^z_{2j},
\nonumber
\\
\hspace{0.3cm}
\varphi_{3}(j)\varphi_{4}(j)
=
\frac{i}{2}
\sigma^x_{2j},
\hspace{0.3cm}
\varphi_{4}(j)\varphi_{1}(j+1)
=
\frac{i}{2}
\sigma^z_{2j}\sigma^z_{2j+1}.
\end{eqnarray}

Let us consider the Fourier transform of $\varphi_{k}(j)$ for each $k=1, 2, 3, 4$ as 
\begin{eqnarray}
\varphi_{k}(j)
=
\frac{1}{\sqrt{N/2}}\sum_{0<p<\pi}(e^{ipj}c_k(p)+e^{-ipj}c_k^\dag(p)),
\label{fouriertr}
\end{eqnarray}
where 
\begin{eqnarray}
\{c_k^\dag(p), c_l(q)\}
=
\delta_{kl}\delta_{pq},
\hspace{0.4cm}
\{c_k^\dag(p), c_l^\dag(q)\}
=
\{c_k(p), c_l(q)\}=0, 
\label{comcpcq}
\end{eqnarray}
and the condition (\ref{bcond}) yields that 
\begin{eqnarray}
p=\frac{l}{N/2}\pi,
\hspace{0.6cm}
 l=
\left\{
\begin{array}{ll}
1, 3, \ldots, \frac{N}{2}-1 & \sigma_1^x\cdots\sigma_N^x=+1,\\
0, 2, \ldots, \frac{N}{2} & \sigma_1^x\cdots\sigma_N^x=-1.
\end{array}
\right.
\end{eqnarray}
In the transformation (\ref{fouriertr}), we will consider the case $l={\rm odd}$; 
when $l={\rm even}$, exceptional treatment is needed for $l=0$ and $N/2$
but the difference between $l={\rm even}$ and odd, 
which comes from the boundary condition (\ref{bcond}), 
finally vanishes in the thermodynamic limit. 
We obtain  
\begin{eqnarray}
\sum_{j=1}^{N/2}
\varphi_{k}(j)\varphi_{k+1}(j)
&=&
\sum_{0<p<\pi}c_{k \:k+1}(p)
\hspace{0.6cm}
(k=1, 2, 3),
\nonumber
\\
\sum_{j=1}^{N/2}
\varphi_{4}(j)\varphi_{1}(j+1)
&=&
\sum_{0<p<\pi}{\tilde c}_{41}(p),
\end{eqnarray}
where 
\begin{eqnarray}
c_{kl}(p)
&=&
c_{k}(p)c_{l}^\dag(p)+c_{k}^\dag(p)c_{l}(p),
\nonumber
\\
{\tilde c}_{kl}(p)
&=&
e^{-ip}c_{k}(p)c_{l}^\dag(p)+e^{ip}c_{k}^\dag(p)c_{l}(p).
\end{eqnarray}
Then the transfer matrices (\ref{TransferMatrix}) are expressed as 
\begin{eqnarray}
V_{1}
&=&
\Big(\frac{e^{K_{1}}}{\cosh K_{1}^*}\Big)^{N/2}
\Big(\frac{e^{K_{2}}}{\cosh K_{2}^*}\Big)^{N/2}
\nonumber
\\
&&
\times \exp\Big[
(-2i)K_{1}^*\sum_{0<p<\pi} c_{12}(p)+(-2i)K_{2}^* \sum_{0<p<\pi}c_{34}(p)
\Big],
\nonumber
\\
V_{2j}
&=&
\exp\Big[
(-2i)L_{j}
\big(
\sum_{0<p<\pi}c_{23}(p)+\sum_{0<p<\pi}{\tilde c}_{41}(p)
\big)
\Big].
\label{TransferMatrix-q}
\end{eqnarray}
The operators $c_{kl}(p)$ and ${\tilde c_{kl}(p)}$ satisfy
\begin{eqnarray}
&&
[c_{kl}(p), c_{lm}(p)]=c_{km}(p),
\hspace{0.4cm}
[c_{kl}(p), {\tilde c_{lm}(p)}]={\tilde c_{km}(p)},
\nonumber
\\
&&
c_{kl}(p)^2={\tilde c}_{kl}(p)^2
\hspace{0.4cm}
c_{kl}(p)^3=-c_{kl}(p),
\hspace{0.4cm}
{\tilde c}_{kl}(p)^3=-{\tilde c}_{kl}(p), 
\end{eqnarray}
where $k$, $l$ and $m$ are all distinct from each other.

When the operator $c$ satisfies $c^3=-c$, we obtain 
\begin{eqnarray}
\exp(xc)
=
{\bf 1}+c\sin x+c^2(1-\cos x),
\label{exp(xc)}
\end{eqnarray}
where ${\bf 1}$ is the unit operator. 
With the use of the relation 
$c_{12}(p)c_{1}^\dag(p)c_{12}(p)=0$, 
we obtain 
\begin{eqnarray}
&&
\exp(xc_{12}(p))c_{1}^\dag(p)\exp(-xc_{12}(p))
\nonumber
\\
&=&
({\bf 1}+c_{12}(p)\sin x+c_{12}(p)^2(1-\cos x))
c_{1}^\dag(p)
({\bf 1}-c_{12}(p)\sin x+c_{12}(p)^2(1-\cos x)).
\nonumber
\\
&=&
c_{1}^\dag(p)\cos x -c_{2}^\dag(p)\sin x .
\label{Ec2dagE}
\end{eqnarray}
Similarly one obtains 
\begin{eqnarray}
\exp(xc_{12}(p))c_{2}^\dag(p)\exp(-xc_{12}(p))
=
c_{1}^\dag(p)\sin x +c_{2}^\dag(p)\cos x .
\label{Ec1dagE}
\end{eqnarray}

The operator $\exp(xc_{12}(p))$ itself acts on a $2^N$-dimensional space. 
However, on the right-hand side of (\ref{Ec2dagE}) and (\ref{Ec1dagE}), 
we find a rotation matrix 
that operates on a $2N$-dimensional space 
spanned by $c_j(p)^\dag$ and $c_j(p)$. 
This relation is parallel to the fact that 
angular momentum generates  spatial rotation. 
Kaufman pointed out that 
when the $2N$-dimensional rotation has eigenvalues 
$\lambda=e^{\pm i\theta_j}$ $(j=1, 2, \ldots, N)$, 
then the $2^N$-dimensional operator has eigenvalues 
\begin{eqnarray}
\exp[\frac{i}{2}(\pm \theta_1\pm \theta_2\cdots \pm \theta_N)].
\end{eqnarray}

Now we are going to evaluate $\lambda$'s. 
The operators $c_j^\dag(p)$ and $c_j(p)=c_j^\dag(-p)$ $(0<p<\pi)$, 
i.e. the operators $c_j^\dag(p)$ $(-\pi<p<\pi)$, 
where $j=1,2,3,4$,
form a complete basis of the $2N$-dimensional operator space. 
Let $R(V_k)$ be the $2N$-dimensional rotation matrix  
corresponding to the $2^N$-dimensional transfer matrix $V_k$, 
where the overall factor 
$\displaystyle 
\Big(\frac{e^{K_{1}}}{\cosh K_{1}^*}\Big)^{N/2}
\Big(\frac{e^{K_{2}}}{\cosh K_{2}^*}\Big)^{N/2}
$
is excluded. 
We find
\begin{eqnarray}
R(V_1)
&=&
\prod_{-\pi<p<\pi}
\left(
\begin{array}{cccc}
\cos x_1 & -\sin x_1 & 0 &0  \\
\sin x_1 & \cos x_1 &  0 &0  \\
0 & 0  & \cos x_2 & -\sin x_2  \\
0 & 0 & \sin x_2 & \cos x_2  
\end{array}
\right),
\nonumber
\\
R(V_{2j})
&=&
\prod_{-\pi<p<\pi}
\left(
\begin{array}{cccc}
\cos y_j & 0 & 0 &e^{ip}\sin y_j  \\
0 & \cos y_j  &  -\sin y_j  &0  \\
0 & \sin y_j  & \cos y_j  & 0  \\
-e^{-ip}\sin y_j  & 0 & 0 & \cos y_j   
\end{array}
\right),
\end{eqnarray}
where $x_1=(-2i)K_1^*$, $x_2=(-2i)K_2^*$, $y_1=(-2i)L_{1}$ and  $y_2=(-2i)L_{2}$. 
We will consider 
the transfer matrix $V=V_{1}V_{22}V_{1}V_{21}$, 
where 
$R(V_{1}V_{22}V_{1}V_{21})=R(V_{1})R(V_{22})R(V_{1})R(V_{21})$. 
The eigenequation of $R(V)$ has the form 
\begin{eqnarray}
F(\lambda)=\lambda^4-B_3(p)\lambda^3+C(p)\lambda^2-B_1(p)\lambda+1=0, 
\label{seceq}
\end{eqnarray}
where we have used the fact that the determinant of the matrix is equal to $1$. 

Here we will show that 
$B_3(p)=B_1(p)$. 
Inverse of the transfer matrix 
\begin{eqnarray}
R(V)
=R(V_{1}(x_1, x_2))R(V_{22}(y_2))R(V_{1}(x_1, x_2))R(V_{21}(y_1))
\end{eqnarray}
is obtained as 
\begin{eqnarray}
R(V)^{-1}
&=&
R(V_{21}(y_1))^{-1}R(V_{1}(x_1, x_2))^{-1}R(V_{22}(y_2))^{-1}R(V_{1}(x_1, x_2))^{-1}
\nonumber
\\
&=&
R(V_{21}(-y_1))R(V_{1}(-x_1, -x_2))R(V_{22}(-y_2))R(V_{1}(-x_1, -x_2)).
\end{eqnarray}
We can replace $-x_j$ and $-y_j$ with $x_j$ and $y_j$, respectively,  
by flipping the spins on one sublattice, 
while keeping the other spins unchanged. 
The partition function is invariant under this transformation, 
and hence the eigenequation of $R(V)^{-1}$ is identical with that of 
\begin{eqnarray}
R({\bar V})
=
R(V_{21}(y_1))R(V_{1}(x_1, x_2))R(V_{22}(y_2))R(V_{1}(x_1, x_2)). 
\end{eqnarray}
Considering the fact that 
${\rm det\:}(RR'-\lambda E)={\rm det\:}(R'R-\lambda E)$, 
where $E$ is the unit matrix, 
the eigenequation of $R({\bar V})$ is identical with that of $R(V)$. 
(This fact can also be understood directly from the structure of the lattice.) 
Hence,  if $\lambda$ satisfies (\ref{seceq}), 
then $1/\lambda$ also satisfies (\ref{seceq}), 
which results in $B_3(p)=B_1(p)\: (=B(p))$. 

Then the equation (\ref{seceq}) can be written as 
\begin{eqnarray}
(\lambda+\frac{1}{\lambda})^2-B(p)(\lambda+\frac{1}{\lambda})+(C(p)-2)=0.
\label{seceq2}
\end{eqnarray}
In some cases, 
this equation can be solved easily (see (\ref{DOns}) and (\ref{DLeeYang})).

One simple expressions of the coefficients $B(p)$ and $C(p)$ are 
\begin{eqnarray}
-B(p)
&=&
4(\cos x_1\cos x_2-\cos p \sin x_1\sin x_2)\sin y_1\sin y_2
\nonumber
\\
&+&
2(\sin^2 x_1-\cos^2 x_1+\sin^2 x_2-\cos^2 x_2)\cos y_1\cos y_2,
\nonumber
\\
B(p)^2-4(C(p)-2)
&=&
8[1-(\sin^2 x_1-\cos^2 x_1)(\sin^2 x_2-\cos^2 x_2)](1-\sin^2 y_1\sin^2 y_2)
\nonumber
\\
&+&
32\cos p\sin x_1\sin x_2
[
\cos x_1\cos x_2(\sin^2 y_1\cos^2 y_2+\cos^2 y_1\sin^2 y_2)
\nonumber
\\
&&
\hspace{3.0cm}
+(\cos^2 x_1+\cos^2 x_2)\sin y_1\cos y_1\sin y_2\cos y_2
]
\nonumber
\\
&+&
32(\sin^2 x_1+\sin^2 x_2)\cos x_1\cos x_2 \sin y_1\cos y_1\sin y_2\cos y_2
\nonumber
\\
&-&
16(\sin^2 x_1\cos^2 x_1+\sin^2 x_2\cos^2 x_2)\cos^2 y_1\cos^2 y_2.
\label{BC}
\end{eqnarray}

Let 
$\lambda_1(p)^2$, $1/\lambda_1(p)^2$, $\lambda_2(p)^2$ and  $1/\lambda_2(p)^2$, 
where we assume $|\lambda_1(p)|>1$, $|\lambda_2(p)|>1$, 
be the solutions of (\ref{seceq}) with fixed $p$. 
The transfer matrix $V$ concerns two layers of the square lattice, 
and hence 
$\lambda_1(p)$, $1/\lambda_1(p)$, $\lambda_2(p)$ and  $1/\lambda_2(p)$ 
are the solutions per layer of the lattice. 
Let $\Lambda_{\rm max}^2$ 
be the maximum eigenvalue of the transfer matrix $V$, 
which means that  
$\Lambda_{\rm max}$ is the maximum eigenvalue of $V$ per layer. 
When the rotation matrix has the eigenvalues 
$\lambda_j(p)^{\pm 2}=e^{\pm 2\gamma_j(p)}$ $(j=1, 2)$, 
the eigenvalues of $V$ are the overall factor times 
$\prod_{-\pi<p<\pi}e^{\frac{1}{2}(\pm 2\gamma_1(p)\pm 2\gamma_2(p))}$. 
Then $\Lambda_{\rm max}^2$ is obtained as   
\begin{eqnarray}
\Lambda_{\rm max}^2
=
\Big(\frac{e^{K_{1}}}{\cosh K_{1}^*}\Big)^{N}
\Big(\frac{e^{K_{2}}}{\cosh K_{2}^*}\Big)^{N}
\prod_{-\pi<p<\pi}\lambda_1(p)\lambda_2(p).
\end{eqnarray}
Then we have
\begin{eqnarray}
\frac{1}{N}\log\Lambda_{\rm max}^2
&=&
\log\Big(\frac{e^{K_{1}}}{\cosh K_{1}^*}\Big)\Big(\frac{e^{K_{2}}}{\cosh K_{2}^*}\Big)
+
\frac{1}{4\pi}
\sum_{-\pi<p<\pi} \log \lambda_1(p)\lambda_2(p)
\:\Delta p,
\end{eqnarray}
where 
we used the fact that 
$\displaystyle \Delta p=\frac{2}{N/2}\pi$, 
which comes from 
$\displaystyle p=\frac{l}{N/2}\pi$ $(l=1, 3, \ldots, \frac{N}{2}-1)$. 
The duality relation yields 
$\displaystyle \frac{e^{K_{j}}}{\cosh K_{j}^*}=(2\sinh 2K_j)^{1/2}$. 
Then the free energy in the thermodynamic limit is obtained as 
\begin{eqnarray}
-\beta f
&=&
\lim_{N\to\infty}\frac{1}{N}\log\Lambda_{\rm max}
\nonumber
\\
&=&
\frac{1}{2}\log(2\sinh 2K_1)^{1/2}(2\sinh 2K_2)^{1/2}
+
\frac{1}{2}\frac{1}{4\pi}
\int_{-\pi}^{\pi}\log\lambda_1(p)\lambda_2(p)\:dp.
\label{f-gen}
\end{eqnarray}
Let us introduce the formula\cite{50Temperley} 
\begin{eqnarray}
\frac{1}{2\pi}
\int_0^{2\pi}
\log\Big[
\frac{1}{e^{2i\omega}}F(-e^{i\omega})
\Big]\:d\omega
=
\log\lambda_1(p)^2\lambda_2(p)^2 
\end{eqnarray}
(note that  $\lambda_1(p)^2$ and $\lambda_2(p)^2$ 
are two solutions of the equation $F(x)=0$). 
In our case 
\begin{eqnarray}
\frac{1}{e^{2i\omega}}F(-e^{i\omega})
&=&
\frac{1}{e^{2i\omega}}
(e^{4i\omega}+B(p)e^{3i\omega}+C(p)e^{2i\omega}+B(p)e^{i\omega}+1)
\nonumber
\\
&=&
4\cos^2\omega+2B(p)\cos\omega+(C(p)-2)
\nonumber
\\
&=&
4\Big(\cos\omega+\frac{1}{4}B(p)\Big)^2-\frac{1}{4}\Big(B(p)^2-4(C(p)-2)\Big).
\label{F}
\end{eqnarray}
The free energy (\ref{f-gen}) is written as 
\begin{eqnarray}
-\beta f
&=&
\frac{1}{4}\log(2\sinh 2K_1)(2\sinh 2K_2)
+
\frac{1}{2}\frac{1}{4\pi}
\int_{-\pi}^{\pi}
\:dp
\:\frac{1}{2}\frac{1}{2\pi}
\int_0^{2\pi}
\:d\omega
\log\Big[
\frac{1}{e^{2i\omega}}F(-e^{i\omega})
\Big],
\nonumber\\
\label{FreeEngGen}
\end{eqnarray}
where the function $F$ is given by (\ref{F}). 
This is the most general formula obtained in this paper.

\section{Onsager's Free Energy}
Let us consider the case 
$K_1=K_2=K$ and $L_1=L_2=L$, 
and therefore we can write 
$x_1=x_2=x$, where $x=(-2i)K^*$, and 
$y_1=y_2=y$, where $y=(-2i)L$. 
In this case, 
the system reduces to the rectangular lattice Ising model 
solved by Onsager.
We find  
\begin{eqnarray}
\frac{1}{4}B(p)
&=&
2\cos^2\frac{p}{2}\sin^2x\sin^2y+2\cos^2x\cos^2y-1,
\nonumber
\\
B(p)^2-4(C(p)-2)
&=&
16^2\cos^2\frac{p}{2}\sin^2x\cos^2x\sin^2y\cos^2y.
\label{DOns}
\end{eqnarray}
Let $B(p)^2-4(C(p)-2)=D(p)^2$. 
Then the free energy is written as 
\begin{eqnarray}
-\beta f
&=&
\frac{1}{2}\log(2\sinh 2K)
\nonumber
\\
&+&
\frac{1}{2}\frac{1}{4\pi}
\frac{1}{2}\frac{1}{2\pi}
\int_{-\pi}^{\pi}
\:dp
\int_0^{2\pi}
\:d\omega
\log
\:4
\Big[
(\cos\omega+\frac{1}{4}B(p))-\frac{1}{4}D(p)
\Big]
\Big[
(\cos\omega+\frac{1}{4}B(p))+\frac{1}{4}D(p)
\Big].
\nonumber
\\
\end{eqnarray}
Here we find 
\begin{eqnarray}
&&
(\cos\omega+\frac{1}{4}B(p))\pm\frac{1}{4}D(p)
\nonumber
\\
&=&
\cos\omega+2\cos^2\frac{p}{2}\sin^2x\sin^2y+2\cos^2x\cos^2y-1
\pm 4\cos\frac{p}{2}\sin x\cos x\sin y\cos y
\nonumber
\\
&=&
2(\cos x\cos y\pm\cos\frac{p}{2}\sin x\sin y)^2-2\sin^2\frac{\omega}{2}
\nonumber
\\
&=&
2
(\cos x\cos y\pm\cos\frac{p}{2}\sin x\sin y-\sin\frac{\omega}{2})
(\cos x\cos y\pm\cos\frac{p}{2}\sin x\sin y+\sin\frac{\omega}{2})
\end{eqnarray}
Let $q=p/2$ and $\theta=\omega/2$, 
and arranging the integrals 
to be an integration over the period of trigonometric functions, 
we find   
\begin{eqnarray}
-\beta f
&=&
\frac{1}{2}\log(2\sinh 2K)+\frac{1}{4}\log 2+\frac{1}{4}\log 2
\nonumber
\\
&+&
\frac{1}{2}\frac{1}{4\pi}
\frac{1}{2}\frac{1}{2\pi}
\int_{-\pi}^{\pi}
\:dp
\int_{-\pi}^{\pi}
\:d\theta
\:4
\log
\Big[
\cos x\cos y-\cos q\sin x\sin y-\cos \theta
\Big].
\end{eqnarray}
Substituting $x=(-2i)K^*$ and $y=(-2i)L$, 
and with the use of the duality relations 
$\cosh 2K^*=\cosh 2K/\sinh 2K$ and 
$\sinh 2K^*\sinh 2K=1$, 
we obtain 
\begin{eqnarray}
-\beta f
=
\log 2
+
\frac{1}{2\pi^2}
\int_{0}^{\pi}
\:dq
\int_0^{\pi}
\:d\theta
\log
\Big[
\cosh 2K\cosh 2L-\sinh 2L\cos q-\sinh 2K\cos \theta
\Big], 
\label{OnsagerF}
\end{eqnarray}
which is the free energy derived by Onsager.\cite{44Onsager}

\section{Frustration and Imaginary Field $i\pi/2$}
Kadanoff and Ceva\cite{71KadanoffCeva} considered the following relation, 
which is valid for the Pauli spin operators, that 
\begin{eqnarray}
e^{\frac{i\pi}{2}(\sigma_i^z+\sigma_j^z)}
=
(-1) \sigma_i^z\sigma_j^z
=
i e^{\frac{i\pi}{2}\sigma_i^z \sigma_j^z}.
\label{BoltzmannFac2}
\end{eqnarray}
This relation yields that 
an imaginary field $i\pi/2$ applied to sites $i$ and $j$ 
results in a shift of the interaction $K_{ij}$ between these two sites, i.e.  
\begin{eqnarray}
K_{ij}\mapsto K_{ij}+\frac{i\pi}{2}.
\label{KKpi2}
\end{eqnarray}
When we consider the dual of this interaction, we find 
\begin{eqnarray}
K_{ij}^*\mapsto (K_{ij}+\frac{i\pi}{2})^*=-K_{ij}^*.
\label{KKpi2dual}
\end{eqnarray}
Thus the imaginary magnetic field at the ends of an interaction, 
results in the shift of the interaction strength by $i\pi/2$, 
and results in the change of sign in its dual lattice. 
The change of sign of one interaction 
results in two frustrated squares, as shown in Fig.1(a).

Consider a product of the left-hand side of (\ref{BoltzmannFac2}) 
\begin{eqnarray}
\prod_{j=1}^{l-1} e^{\frac{i\pi}{2}(\sigma_{j}^z+\sigma_{j+1}^z)}
=
(-1)^{l-2}e^{\frac{i\pi}{2}\sigma_{1}^z}e^{\frac{i\pi}{2}\sigma_{l}^z}
\hspace{1.2cm}
(l\geq 2)
\label{ProdCipi}
\end{eqnarray}
Two operators $\sigma_{1}^z$ and $\sigma_{l}^z$ are connected by a path $C$. 
The left-hand side of (\ref{ProdCipi}) is the product of  
$e^{\frac{i\pi}{2}(\sigma_{j}^z+\sigma_{j+1}^z)}$ along $C$. 
The factor $(-1)^{l-2}$ equals $-1$ 
if $\sigma_{1}^z$ and $\sigma_{l}^z$ are on the same sublattice, 
and $+1$ if they are on different sublattices.
The left-hand side results in the shift (\ref{KKpi2}) of interactions on the path $C$, 
while the right-hand side is the Boltzmann factor 
coming from the imaginary field $i\pi/2$ applied at the two ends of the path. 
Fig.1(b) and 1(c) illustrate this situation. 

We find that a magnetic field $i\pi/2$ applied at both ends of the path $C$ 
on the right-hand side of (\ref{ProdCipi}) 
is equivalent to the shift of interactions 
$K_{ij}\mapsto K_{ij}+i\pi/2$ along $C$, 
as can be seen from (\ref{BoltzmannFac2}) and the left-hand side of (\ref{ProdCipi}). 
When we consider its dual lattice, 
we find the interactions $-K_{ij}^*$ along $C$, 
and we also find the frustrated squares at the two ends of the path $C$, 
as shown in Fig.1(a)-(c). 
Hence, we find that an imaginary magnetic field at the ends of the path $C$ 
corresponds to the two frustrated squares at the ends of the path $C$ in the dual lattice.

Let us consider a square lattice with periodic boundary conditions in two directions. 
In this case, frustrated squares appear on both sides of 
$-K_{ij}^*$. 
If two frustrations overlap in a single square, the frustration disappears. 
Therefore, the total number of frustrated squares is even. 

Assume that the number of sites subjected to the imaginary magnetic field is even.
Let the coordinates of the sites under the field $i\pi/2$ be 
\begin{eqnarray*}
&&
(1, \tau^{(1)}_{1}) 
\hspace{0.2cm}
(1, \tau^{(1)}_{2}) 
\hspace{0.2cm}
\ldots
\hspace{0.2cm}
(1, \tau^{(1)}_{m_1}) 
\\
&&
(2, \tau^{(2)}_{1}) 
\hspace{0.2cm}
(2, \tau^{(2)}_{2}) 
\hspace{0.2cm}
\ldots
\hspace{0.2cm}
(2, \tau^{(2)}_{m_2}) 
\\
&&
\hspace{0.8cm}
\vdots
\\
&&
(N, \tau^{(N)}_{1}) 
\hspace{0.2cm}
(N, \tau^{(N)}_{2}) 
\hspace{0.2cm}
\ldots
\hspace{0.2cm}
(N, \tau^{(N)}_{m_N}), 
\end{eqnarray*}
where $\tau^{(k)}_{j}<\tau^{(k)}_{j+1}$. 
First, let us pair the two sites $(1, \tau^{(1)}_{1})$ and $(1, \tau^{(1)}_{2})$, 
and connect them via a path $C_{11}$. 
Next, pair the two sites $(1, \tau^{(1)}_{3})$ and $(1, \tau^{(1)}_{4})$, 
and connect them via a path $C_{13}$. 
Generally, pair the two sites $(1, \tau^{(1)}_{2j-1})$ and $(1, \tau^{(1)}_{2j})$, 
and connect them via a path $C_{1\, 2j-1}$. 
When all the sites in the first column under the magnetic field $i\pi/2$ are classified into pairs, 
next pair the two sites $(2, \tau^{(2)}_{1})$ and $(2, \tau^{(2)}_{2})$, 
and connect them via a path $C_{21}$. 
When the site $(1, \tau^{(1)}_{m_1})$ remains unpaired, 
pair the two sites $(1, \tau^{(1)}_{m_1})$ and $(2, \tau^{(2)}_{m})$ 
where $\tau^{(2)}_{m}$ is the smallest $\tau$ 
that satisfies $\tau^{(1)}_{m_1}\leq \tau^{(2)}_{m}$, 
and where the site $(2, \tau^{(2)}_{1})$ is regarded as $(2, \tau^{(2)}_{m_2+1})$  
because of the periodic boundary condition. 
Next, pair the two sites $(2, \tau^{(2)}_{m+1})$ and $(2, \tau^{(2)}_{m+2})$, 
and connect them via a path $C_{2\: m+1}$. 
All the paths $C_{ij}$ necessary for this pairing
are classified into two types shown in Fig.1(a)-(b) and 1(c). 
By repeating this process, 
all sites subjected to the magnetic field $i\pi/2$ 
are divided into pairs, and the path $C_{ij}$ do not intersect with each other. 
Frustrated squares appear at both ends of the path, 
and there is a one-to-one correspondence between $i\pi/2$ and the frustrated squares.

It is also straightforward to convince that the same result is obtained 
when the path passes through a site subjected to a magnetic field $i\pi/2$, 
or when two paths intersect. 

The method for creating pairs and the method for obtaining paths are not unique. 
Therefore, the arrangement of 
$-K_{ij}^*$ bonds is not unique for a given distribution of  $i\pi/2$. 
However, each imaginary magnetic field $i\pi/2$ 
always corresponds to a single frustrated square. 

When we consider a complete dimer covering (or Domino tiling) of a lattice, 
and add the imaginary interaction $i\pi/2$ for each dimer coupling, 
this is equivalent to applying an imaginary field $i\pi/2$ to all sites in the system. 

One possible distribution of interactions $K+i\pi/2$, 
which yields the Lee-Yang partition function $Z_{LY}$, 
is shown in Fig.2(a). 
When we consider its dual lattice shown in Fig.2(b), 
the partition function $Z_{FF}(K)$ 
of this dual fully frustrated system satisfies
\cite{91Suzuki2}
\begin{eqnarray}
\frac{1}{MN}\log Z_{FF}(K)
=
\frac{1}{MN}\log Z_{LY}(K^*)
+\log(\sinh 2K)-\frac{i\pi}{2},
\label{LYdualVillain}
\end{eqnarray}
where 
the logarithms are assumed to take their principal values and 
$MN$ is the number of sites.

\section{Lee-Yang's Free Energy}
Here $Z_{FF}(K)$, 
and its thermodynamic limit $f_{FF}$,
can be obtained from our result (\ref{FreeEngGen}). 
Let us consider the case 
$K_1=K_2=K$, 
and therefore we can write 
$x_1=x_2=x$, where $x=(-2i)K^*$. 
In this case we find 
\begin{eqnarray}
\frac{1}{4}B(p)
&=&
2\cos^2\frac{p}{2}\sin^2x\sin y_1\sin y_2
-\sin y_1\sin y_2- (\sin^2x-\cos^2x)\cos y_1\cos y_2
\nonumber
\\
B(p)^2-4(C(p)-2)
&=&
16\times 4\cos^2\frac{p}{2}\sin^2x\cos^2x(\sin y_1\cos y_2+\cos y_1\sin y_2)^2.
\end{eqnarray}
Especially when 
$L_1=L$, $L_2=-L$, thus $y_1=y$, $y_2=-y$, where $y=(-2i)L$,
and we find 
\begin{eqnarray}
\frac{1}{4}B(p)
&=&
-2\cos^2\frac{p}{2}\sin^2x\sin^2 y
+\sin^2 y-(\sin^2x-\cos^2x)\cos^2 y 
\nonumber
\\
B(p)^2-4(C(p)-2)
&=&
16\times 4\cos^2\frac{p}{2}\sin^2x\cos^2x(\sin y\cos (-y)+\cos y\sin (-y))^2
\nonumber
\\
&=&
0
\label{DLeeYang}
\end{eqnarray}
Then we obtain 
\begin{eqnarray}
\cos\omega+\frac{1}{4}B(p)
&=&
-2\cos^2\frac{p}{2}\sin^2x\sin^2y
-2\sin^2x\cos^2y+1+\cos\omega
\nonumber
\\
&=&
2(1-\sin^2\frac{\omega}{2}-\sin^2x\cos^2y-\cos^2\frac{p}{2}\sin^2x\sin^2y).
\end{eqnarray}
The free energy is written as 
\begin{eqnarray}
-\beta f
&=&
\frac{1}{2}\log(2\sinh 2K)+\frac{1}{4}\log 2
\nonumber
\\
&+&
\frac{1}{2}
\frac{1}{4\pi}
\int_{-\pi}^{\pi}
\:dp\:
\frac{1}{2}
\frac{1}{2\pi}
\int_0^{2\pi}
\:d\omega
\:2
\log
2
\Big[
1-\sin^2\frac{\omega}{2}-\sin^2x\cos^2y-\cos^2\frac{p}{2}\sin^2x\sin^2y
\Big].\hspace{0.4cm}
\nonumber\\
\label{FreeEngFig2b}
\end{eqnarray}
Let $q=p/2$ and $\theta=\omega/2$. 
Arranging the integrals 
to be an integration over the period of trigonometric functions, 
substituting $x=(-2i)K^*$ and $y=(-2i)L$, 
and with the use of the duality relation 
$\sinh 2K^*\sinh 2K=1$, 
then (\ref{FreeEngFig2b}) can be expressed as 
\begin{eqnarray}
-\beta f
=
\log 2
+
\frac{1}{2\pi^2}
\int_{0}^{\pi/2}
\:dq
\int_0^{\pi}
\:d\theta
\log
\Big(
\sinh^2 2K+\cosh^2 2L
-\sinh^2 2K\sin^2 \theta-\sinh^2 2L\sin^2 q
\Big).
\label{KKL-L}
\end{eqnarray}
This free energy was already obtained in ref.29. 
Let $K=L$, 
then we find $-\beta f_{FF}$. 
Introducing $u=e^{-2K^*}$, 
and using (\ref{LYdualVillain}), 
we find 
\begin{eqnarray}
-\beta f_{LY}
=
\frac{i\pi}{2}+2K^*
+
\frac{1}{4\pi^2}
\int_{0}^{\pi}
\:dq
\int_0^{\pi}
\:d\theta\:
\log(1-u^2)^2
\Big(
1+u^2(6-4\cos^2 q-4\cos^2\theta)+u^4
\Big).
\label{LeeYangF}
\end{eqnarray}
This is the free energy 
obtained by Lee and Yang.\cite{52LeeYang}

\section{Conclusion}
In this paper, 
the free energy of a square lattice Ising model 
with interactions alternating in two directions is derived. 
The model is defined in (\ref{Hamiltonian}). 
The free energy is summarized in (\ref{BC}), (\ref{F}) and (\ref{FreeEngGen}). 

This free energy includes the Onsager's free energy (\ref{OnsagerF}) as a special case, 
and also includes the Lee-Yang's result (\ref{LeeYangF}) as a special case. 

It is derived that 
each imaginary field $i\pi/2$ corresponds to a frustrated plaquette in its dual lattice, 
as summarized in Fig.1. 

Distributions of imaginary bonds determine dimer coverings of the lattice, 
and complete dimer coverings yield the Lee-Yang free energy.

It can also be said that 
a distribution of the frustration 
corresponds to a distribution of the imaginary field $i\pi/2$. 
Thus, models with the same distribution of the frustration 
are equivalent to the same imaginary field model, 
and thus equivalent to each other. 

Finally it should be noted that 
systems with $K_1>0$, $K_2>0$, $L_1>0$, and $L_2<0$ in the dual lattice, 
the free energies of which can be obtained from the main formula (\ref{FreeEngGen}), 
correspond to generalized Lee-Yang systems with four different couplings, 
as can be seen from (\ref{KKpi2}), (\ref{KKpi2dual}) and Fig.2.

\section*{Acknowledgements}
This work was supported by JSPS KAKENHI Grant Number 25K07166.
\\

\includegraphics{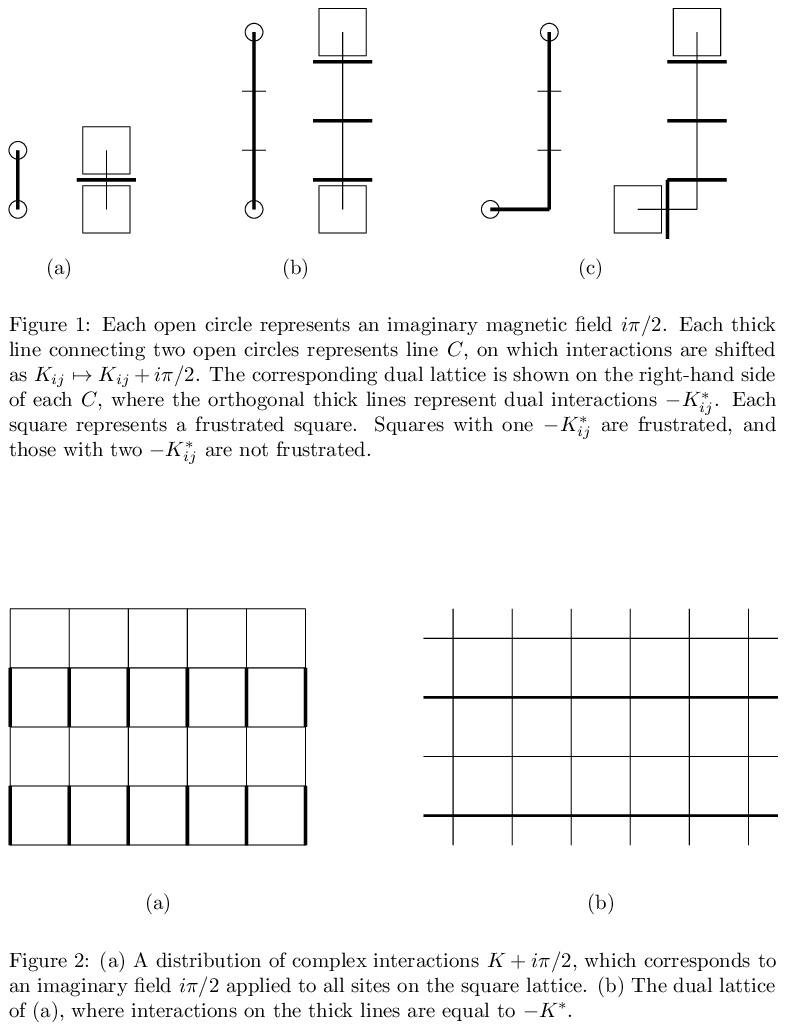}

\end{document}